\documentclass[11pt,a4paper,english,american]{article}
\pdfoutput=1
\usepackage{lmodern}

\usepackage[T1]{fontenc}
\usepackage[latin9]{inputenc}
\usepackage{amsmath}
\usepackage{amssymb}
\usepackage{esint}

\RequirePackage[colorlinks=true
,urlcolor=blue
,anchorcolor=blue
,citecolor=blue
,filecolor=blue
,linkcolor=blue
,menucolor=blue
,linktocpage=true
,pdfproducer=medialab
]{hyperref}

\makeatletter

\usepackage{jcappub}
\numberwithin{equation}{section}

\pdfoutput=1
\renewcommand\[{\begin{equation}}
\renewcommand\]{\end{equation}}

\AtBeginDocument{
  
}

\makeatother

\usepackage{babel}

\begin{document}

\title{The Phantom of the New Oscillatory Cosmological Phase}

\subheader{YITP-16-}

\author[a]{Damien A. Easson}

\author[b,c]{and Alexander Vikman}

\affiliation[a]{Department of Physics \& Beyond Center for Fundamental Concepts
in Science,\\
Arizona State University, Tempe, AZ 85287-1504, USA\\
}

\affiliation[b]{Institute of Physics, the Academy of Sciences of the Czech Republic,
\\
Na Slovance 2, 182 21 Prague 8, Czech Republic\\
}

\affiliation[c]{Yukawa Institute for Theoretical Physics, Kyoto University, Kyoto
606-8502, Japan}

\emailAdd{easson@asu.edu}

\emailAdd{vikman@fzu.cz}

\abstract{We study a recently\emph{ }proposed new cosmological phase where
a scalar field moves periodically in an expanding spatially-flat Friedmann
universe. This phase corresponds to a limiting cycle of the equations
of motion and can be considered as a cosmological realization of a
``time-crystal''. We show that this phase is only possible, provided
the Null Energy Condition is violated and the so-called Phantom divide
is crossed. We prove that in general k-essence models: i) this crossing
causes infinite growth of quantum perturbations on short scales, and
ii) exactly periodic solutions are only possible, provided the limiting
cycle encircles a singularity in the phase plane. The configurations
neighboring this singular curve in the phase space are linearly unstable
on one side of the curve and superluminal on the other side. Moreover,
the increment of the instability is infinitely growing for each mode
by approaching the singularity, while for the configurations on the
other side, the sound speed is growing without limit. We illustrate
our general results by analytical and numerical studies of a particular
class of such k-essence models. }

\maketitle

\section{Introduction}

Usually the appearance of a periodic structure in space is attributed
to formation of ordinary crystals. By analogy, the authors of \cite{Shapere:2012nq}
suggested to use the name ``time-crystals'' for the systems with
vacuum solutions which are exactly periodic in time. In cosmology
such vacuum solutions can be interesting in context of Dark Energy
and early stages of Inflation. On the quantum level this idea was
further elaborated in \cite{Wilczek:2012jt}. However, a cosmological
expansion usually works as a friction and dissipates energy, so that
an exactly periodic motion is impossible. Indeed, a usual system will
finally approach an equilibrium static configuration with minimal
energy density. However, in some non-canonical cases, the lowest possible
available configuration can still be non-static and correspond to
a motion. In particular, this is the case for the ghost condensate
\cite{ArkaniHamed:2003uy}. This is also the basis for the original
k-inflation \cite{ArmendarizPicon:1999rj} attractor. Thus the solution
with the lowest energy level can spontaneously break the time-translation
invariance. Due to the shift-symmetry of the ghost condensate, this
spontaneous symmetry breaking still results there in the time-independent
and Lorentz-invariant energy momentum tensor (EMT). The latter corresponds
to the normal de Sitter vacuum with some cosmological constant. It
is rather interesting to consider wether one can spontaneously break
time-translational invariance even on the level of the EMT so that
the vacuum configuration is different from the de Sitter spacetime.
Here we mean that this non-trivial solution should be valid for very
long times up to the asymptotic future. In particular, if the motion
in this vacuum state is periodic, with the period $\mathcal{T}$,
then the continuous time translation invariance, $t\rightarrow t+c$
with $c$ real, is only broken to the level of a discrete subgroup
$t\rightarrow t+n\mathcal{T}$, where $n$ is an integer. The recent
work \cite{Bains:2015gpv} proposed an interesting new phase of cosmological
matter with these properties. In this phase a scalar field periodically
moves in the constantly expanding universe. This oscillatory phase
was realized by a non-canonical scalar field theory of the k-essence
type \cite{ArmendarizPicon:1999rj,ArmendarizPicon:2000ah,ArmendarizPicon:2000dh,Chiba:1999ka},
see the next section for details. A similar idea of an oscillating
dark energy was studied in e.g. \cite{Feng:2004ff,Nojiri:2006ww}.
In this paper we discuss weather it is possible to realize a k-essence
time-crystal in a cosmological setup.

\section{NEC violation is needed for a limiting cycle }

A presence of exactly periodic motion with the period $\mathcal{T}$,
in particular, requires that the energy density is periodic $\varepsilon\left(t+\mathcal{T}\right)=\varepsilon\left(t\right)$.
If at $t$ the energy density was decreasing, then somewhere between
$t$ and $t+\mathcal{T}$ the energy density should start to increase
to compensate this reduction. And vice versa: for the originally increasing
energy density there should be a moment between $t$ and $t+\mathcal{T}$
where the energy density should start to decrease. Thus anyway during
each cycle there are time intervals on which the energy density is
growing and on which it is decreasing. Further, the conservation of
energy in a Friedmann universe requires 
\begin{equation}
\dot{\varepsilon}=-3H\left(\varepsilon+p\right)\,,\label{eq:energy_conservation}
\end{equation}

where $H$ is the Hubble parameter and $p$ is the effective pressure
of the oscillatory matter. Hence, in an expanding universe (with $H>0$)
the energy density can grow, only if the effective enthalpy density
$\varepsilon+p$ is negative. The negative sign of $\varepsilon+p$
implies a violations of the Null Energy Condition (NEC) which states
that for all null vectors $n^{\mu}$ the energy-momentum tensor should
satisfy $T_{\mu\nu}n^{\mu}n^{\nu}\geq0$. In turn, a violation of
the NEC necessarily implies that the Hamiltonian density of the system
is unbounded from below - for each constant there are such local values
of the initial data that the energy density is more negative than
this constant, see \cite{Sawicki:2012pz}. In other words a violation
of NEC implies that the system has to possess configurations with
arbitrary negative energy densities. For a recent discussion of NEC
violation see e.g. \cite{Rubakov:2014jja}. Clearly, in a collapsing
universe, a decrease in energy density would require an NEC violation
as well. In cosmology, matter which violates NEC is often referred
to as Phantom \cite{Caldwell:1999ew}. 

On the other hand, the time derivative of the energy density can change
the sign, provided $H$ changes sign. In this case the spatially-flat
Friedmann universe has to be able to evolve from expansion to contraction
and then from contraction to expansion. The latter transition corresponds
to a bounce. It is well known that a bounce of a spatially-flat Friedmann
universe requires a violation of NEC. Indeed, the transition from
contraction to expansion at time $t_{b}$ requires $H\left(t_{b}\right)=\dot{a}/a=0$.
Using the Friedmann equations 
\begin{equation}
H^{2}=\frac{1}{3}\varepsilon\,,\label{eq:Friedmann_1}
\end{equation}
and 
\begin{equation}
\dot{H}=-\frac{1}{2}\left(\varepsilon+p\right)\,,\label{eq:Friedmann_2}
\end{equation}
we obtain that at transition from contraction with $H<0$ to expansion
with $H>0$ one has $\dot{H}>0$ and therefore $\varepsilon+p<0$
so that NEC is violated again. Hence we conclude that a violation
of NEC is necessary to realize an exactly periodic oscillatory cosmological
phase. Moreover, due to the periodicity the system should be able
to evolve through the so-called Phantom divide - the border of the
NEC-violating region. This border corresponds to the equation of state
parameter $w=p/\varepsilon=-1$. The dynamical transition through
$w=-1$ should cyclicly happen in both directions from above and from
below. 

It is well known that systems of k-essence type cannot dynamically
violate NEC and cross the Phantom divide \cite{Vikman:2004dc}, see
also \cite{Hu:2004kh,Caldwell:2005ai,Xia:2007km} and for the review
\cite{Cai:2009zp,Nesseris:2006er}. Note that the pathologies associated
with the crossing of the Phantom divide are more severe than those
problems \cite{Buniy:2005vh,Buniy:2006xf,Dubovsky:2005xd,Abramo:2005be}
arising just due to the violation of NEC. Below in the next section
(\ref{sec:Refreshing-k-essence-and}) we will refresh some basic facts
about k-essence and provide simple arguments against a smooth transition
of a k-essence field through the Phantom divide. 

Finally it is useful to consider the average of $w+1$ over some time
interval $T=t_{f}-t_{i}$. Using (\ref{eq:energy_conservation}) and
(\ref{eq:Friedmann_1}) under the assumption that $\varepsilon>0$
between $t_{i}$ and $t_{f}$ we obtain 
\[
\left\langle w\left(t\right)+1\right\rangle _{T}=\frac{1}{T}\int_{t_{i}}^{t_{f}}dt\left(1+w\left(t\right)\right)=-\frac{1}{T}\int_{t_{i}}^{t_{f}}dt\frac{\dot{\varepsilon}}{\varepsilon\sqrt{3\varepsilon}}=\frac{1}{T}\sqrt{\frac{4}{3}}\left(\frac{1}{\sqrt{\varepsilon\left(t_{f}\right)}}-\frac{1}{\sqrt{\varepsilon\left(t_{i}\right)}}\right)\,.
\]
If the time interval $T$ is a multiple of the period of oscillations
$\mathcal{T}$, then $\left\langle w\left(t\right)+1\right\rangle _{T}=0$.
Thus such an oscillatory stage can be interesting to model inflation
and dark energy.

\section{Refreshing k-essence and why cannot it cross the Phantom divide\label{sec:Refreshing-k-essence-and} }

Here we will collect and discuss basic mostly well known facts about
k-essence which is a noncanonical, minimally coupled to gravity scalar
field whose dynamics is described by the action 

\begin{equation}
S=\int d^{4}x\,\sqrt{-g}\,p\left(\varphi,X\right)\,,\label{eq:action_general}
\end{equation}
where 
\begin{equation}
X=\frac{1}{2}g^{\mu\nu}\varphi_{,\mu}\varphi_{,\nu}\,.\label{eq:X}
\end{equation}
In cosmological applications the field is only slightly inhomogeneous
and anisotropic so that $\varphi_{,\mu}$ is timelike, hence throughout
the paper we assume that $X>0$, as our signature convention $\left(+,-,-,-\right)$.
This class of theories was introduced in \cite{ArmendarizPicon:1999rj}
(see also \cite{ArmendarizPicon:2000dh,ArmendarizPicon:2000ah,Chiba:1999ka}).
The corresponding EMT is 
\begin{equation}
T_{\mu\nu}=p_{,X}\varphi_{,\mu}\varphi_{,\nu}-pg_{\mu\nu}\,,\label{eq:EMT_General}
\end{equation}
for every timelike $\varphi_{,\mu}$ this EMT takes a form of a perfect
fluid 
\begin{equation}
T_{\mu\nu}=\left(\varepsilon+p\right)u_{\mu}u_{\nu}-pg_{\mu\nu}\,,\label{eq:EMT_Fluid}
\end{equation}
with the four velocity 
\begin{equation}
u_{\mu}=\frac{\varphi_{,\mu}}{\sqrt{2X}}\,,\label{eq:Velocity}
\end{equation}
pressure $p$ and the energy density 
\begin{equation}
\varepsilon=2Xp_{,X}-p.\label{eq:Energy_density}
\end{equation}
Hence for all null vectors $n^{\mu}$ we have 
\begin{equation}
T_{\mu\nu}n^{\mu}n^{\nu}=p_{,X}\left(\varphi_{,\mu}n^{\mu}\right)^{2}=\frac{\varepsilon+p}{2X}\left(\varphi_{,\mu}n^{\mu}\right)^{2}\,.\label{eq:NEC_k}
\end{equation}
The Null Energy Condition requires $T_{\mu\nu}n^{\mu}n^{\nu}\geq0$
or $p_{,X}\geq0$. Thus to change the sign of $T_{\mu\nu}n^{\mu}n^{\nu}$
the system has to change the sign of $p_{,X}$ or for time-like derivatives
to change the sign of the enthalpy density $\varepsilon+p$. In particular,
if the transition is smooth, then at the Phantom divide $p_{,X}=0$. 

Around any time-like, $X>0$, background (in particular around cosmological
backgrounds \cite{Garriga:1999vw}) the perturbations propagate with
the speed given by 
\begin{equation}
c_{\text{S}}^{2}=\left(\frac{\partial p}{\partial\varepsilon}\right)_{\varphi}=\frac{p_{,X}}{\varepsilon_{,X}}\,.\label{eq:sound}
\end{equation}
 The small perturbations of k-essence around any background propagate
in an effective (contravariant) metric \cite{Babichev:2007dw} which
is conformally equivalent to 
\[
G_{\mu\nu}=p_{,X}g_{\mu\nu}+p_{,XX}\varphi_{,\mu}\varphi_{,\nu}\,,
\]
see also \cite{Babichev:2007dw,ArmendarizPicon:2005nz,Bruneton:2006gf,Bruneton:2007si},
see also \cite{Moncrief:1980,Bilic:1999sq} for the relativistic acoustic
geometry in closely related irrotational perfect fluids. The equation
of motion is 
\begin{equation}
G^{\mu\nu}\nabla_{\mu}\nabla_{\nu}\varphi+\varepsilon_{,\varphi}=0\,.\label{eq:covariant_equation_of_motion}
\end{equation}
The positivity of the right hand side of the expression (\ref{eq:sound})
guaranties that this equation of motion is a hyperbolic quasilinear
PDE. Note that hyperbolicity guaranties that Cauchy problem is well
posed and evolution is predictable till some moment of time. However,
it was demonstrated that breakdown of predictability due to the formation
of caustics is a rather generic phenomenon, see recent discussion
in \cite{Babichev:2016hys,Mukohyama:2016ipl}. In cosmology the equation
of motion reduces to 
\begin{equation}
\varepsilon_{,X}\ddot{\varphi}+3H\dot{\varphi}p_{,X}+\varepsilon_{,\varphi}=0\,,\label{eq:equation_of_motion_cosmology}
\end{equation}
where $H$ is given by the first Friedmann equation (\ref{eq:Friedmann_1})
with the energy density (\ref{eq:Energy_density}). Instead of $\varphi$
and $\dot{\varphi}$ one can use $\varepsilon$ and $p$ as independent
variables. In this case the Jacobian is $J=p_{,\varphi}\varepsilon_{,\dot{\varphi}}-\varepsilon_{,\varphi}p_{,\dot{\varphi}}$.
The dynamical equation on energy density is (\ref{eq:energy_conservation})
with the Hubble parameter given by the first Friedmann equation (\ref{eq:Friedmann_1}).
While the dynamical equation for the pressure is 
\begin{equation}
\dot{p}=-\frac{1}{\varepsilon_{,X}}\left(6XHp_{,X}^{2}+p_{,\varphi}\varepsilon_{,\dot{\varphi}}-\varepsilon_{,\varphi}p_{,\dot{\varphi}}\right)\,,\label{eq:derivative_of_pressure}
\end{equation}
where the Hubble parameter is expressed through the energy density
by the first Friedmann equation (\ref{eq:Friedmann_1}) and all other
quantities should be expressed through $\left(\varepsilon,p\right)$
using implicit function theorem. Unfortunately this description is
only valid outside of the $\dot{\varphi}=0$ line. Indeed, the Jacobian
can be written as 
\[
J=p_{,\varphi}\varepsilon_{,\dot{\varphi}}-\varepsilon_{,\varphi}p_{,\dot{\varphi}}=\dot{\varphi}\left(p_{,\varphi}\varepsilon_{,X}-\varepsilon_{,\varphi}p_{,X}\right)\,,
\]
where the right hand side is vanishing on the $\dot{\varphi}=0$ line.
However, if a limiting cycle exists, it has to cross this line. Thus
the Jacobian is not sign-definite and it is not clear whether the
expression in the brackets of (\ref{eq:derivative_of_pressure}) is
sign-definite. For a shift-symmetric case (with the symmetry $\varphi\rightarrow\varphi+c$)
the pressure evolves due to the (\ref{eq:energy_conservation}). In
this case a limiting cycle is not possible in an expanding universe,
because $\dot{\varphi}p_{,X}\propto a^{-3}$ so that $\dot{\varphi}p_{,X}$
always decreasing. 

Further it is useful to use the energy conservation (\ref{eq:energy_conservation})
along with the (\ref{eq:Friedmann_1}) and the definition of energy
density (\ref{eq:Energy_density}) to obtain 
\[
\sqrt{\frac{4}{3}\varepsilon_{f}}-\sqrt{\frac{4}{3}\varepsilon_{i}}=\int_{\varepsilon_{i}}^{\varepsilon_{f}}\frac{d\varepsilon}{\sqrt{3\varepsilon}}=-\int_{t_{i}}^{t_{f}}dt\dot{\varphi}p_{,\dot{\varphi}}=-\int_{\varphi_{i}}^{\varphi_{f}}d\varphi\,p_{,\dot{\varphi}}\,.
\]

\subsection{Momentum invertibility and strong superluminality }

Further it is illuminating to look at the canonical formulation of
the dynamics. It is particularly useful as the authors of \cite{Bains:2015gpv}
are interested in degenerate Hamiltonians which are not smooth functions
of canonical momenta. For smooth Lagrangians this happens when the
field velocity cannot be uniquely expressed through the canonical
momentum. We will proceed using the ADM formalism, see \cite{Arnowitt:1962hi}
and \cite{Poisson}. In a spacetime foliation generated by a time-like
congruence with a tangent vector $t^{\mu}$ we can represent the metric
as 
\[
ds^{2}=N^{2}dt^{2}-\ell_{ik}\left(dx^{i}+N^{i}dt\right)\left(dx^{k}+N^{k}dt\right)\,.
\]
The unit normal to the hypersurface of constant $t$ is 
\[
U_{\mu}=N\partial_{\mu}t\,.
\]
The relative three velocity $v$ of $U_{\mu}$ with respect to the
k-essence fluid velocity $u^{\mu}$ (given by (\ref{eq:Velocity}))
can be found from 
\begin{equation}
\frac{1}{\sqrt{1-v^{2}}}=U_{\mu}u^{\mu}=Nu^{t}=\frac{1}{N\sqrt{2X}}\left(\dot{\varphi}-N^{i}\varphi_{,i}\right)\,,\label{eq:relative_velocity}
\end{equation}
so that after some algebra 
\begin{equation}
v^{2}=\frac{N^{2}\ell^{ik}\varphi_{,i}\varphi_{,k}}{\left(\dot{\varphi}-N^{i}\varphi_{,i}\right)^{2}}\,,\label{eq:relative_v}
\end{equation}
where we have used the standard ADM results $g^{ik}=-\ell^{ik}+N^{i}N^{k}/N^{2}$,
$g^{ti}=-N^{i}/N^{2}$ and $g^{tt}=1/N^{2}$, $\ell^{ik}\ell_{km}=\delta_{m}^{i}$. 

The canonical momentum is defined as 
\begin{equation}
P=\frac{\partial}{\partial\dot{\varphi}}\left(\sqrt{-g}p\right)=\sqrt{-g}p_{,X}\frac{1}{N^{2}}\left(\dot{\varphi}-N^{i}\varphi_{,i}\right)\,.\label{eq:momentum}
\end{equation}
Hence, the canonical momentum is always vanishing at the crossing
of the Phantom divide $p_{,X}=0$. The velocity $\dot{\varphi}$ can
be found from the canonical momentum provided
\begin{equation}
\frac{\partial P}{\partial\dot{\varphi}}\neq0\,.\label{eq:solvability_momentum}
\end{equation}
In cosmology this momentum is $P=a^{3}\dot{\varphi}p_{,X}$ and cannot
be \emph{locally} expressed exclusively through $\varphi$ and $\dot{\varphi}$
as $a\left(t\right)=\exp\left(\int^{t}dt'\sqrt{\varepsilon\left(\varphi\left(t'\right),\dot{\varphi}\left(t'\right)\right)/3}\right)$.
Using the definitions of the relative velocity (\ref{eq:relative_v})
and the sound speed (\ref{eq:sound}) this invertibility condition
(for a given scale factor $a$, before solving the Hamiltonian constraint
- the first Friedmann equation (\ref{eq:Friedmann_1})) can be written
in the form 
\begin{equation}
\frac{\partial P}{\partial\dot{\varphi}}=\sqrt{-g}\left(\left(u^{t}\right)^{2}2Xp_{,XX}+p_{,X}g^{tt}\right)=\sqrt{-g}G^{tt}=\frac{\varepsilon_{,X}}{N}\sqrt{\ell}\left(\frac{1-v^{2}c_{\text{S}}^{2}}{1-v^{2}}\right)\,.\label{eq:invertability}
\end{equation}
In particular, one can chose such a foliation that $\varphi$ is constant
on the equal-time hypersurface, so that $\varphi\left(t\right)$.
In this foliation $v=0$ and the invertibility of the momentum requires
that $\varepsilon_{,X}\neq0$. Note that cusp Hamiltonian
and non-invertibility of momentum are among the desirable features
of the ``time crystals'' and one of the requirements imposed in
\cite{Bains:2015gpv} on a general system with an oscillatory attractor
as a ground state. Indeed, the ground state implies a minimum of the
Hamiltonian. But if all first derivatives of the Hamiltonian are vanishing,
there is no motion and an oscillatory configuration cannot be a ground
state. Clearly for systems with cusp Hamiltonians this argument does
not work. 

However, the homogeneous and isotropic cosmological dynamics of a
k-essence scalar field driving the expansion of the universe cannot
be brought to this simple two dimensional canonical form. Below, in
section \ref{sec:General-phase-space}, using standard results from
the theory of ODE, we prove that the changing of sign of $\varepsilon_{,X}$
is a necessary condition for the existence of cosmological limiting
cycles. Hence one has to require that on some configurations $\varepsilon_{,X}=0$.
But, for a generic k-essence Lagrangian, $\varepsilon_{,X}=0$ does
not imply $p_{,X}=0$. Therefore, for $X>0$ and a finite and non-vanishing
$p_{,X}$ the singular surface where $\varepsilon_{,X}=0$ corresponds
to the divergent speed of sound for the perturbations $c_{\text{S}}^{2}\rightarrow\infty$.
Moreover, if $\varepsilon_{,X}$ changes sign (as required for the
existence of a limiting cycle) it implies that on one side of the
singular hypersurface there is a region of extreme superluminality,
while the configurations on the other side suffer from infinitely
strong gradient instabilities. On these latter configurations $c_{\text{S}}^{2}<0$
and the equation of motion (\ref{eq:covariant_equation_of_motion})
becomes an elliptic quasilinear PDE. Thus generically these ``time
crystals'' k-essence systems require existence of configurations
with an infinitely strong superluminality which are neighboring configurations
with infinitely strong gradient instabilities. 

As it follows from (\ref{eq:derivative_of_pressure}) the pressure
has an infinite time derivative at the singularity. Thus the time
derivative of the Ricci scalar $R$ blows up as well. An effective
action in gravity should contain the term $\left(\partial R\right)^{2}$
which blows up in this case. This is another way to see that EFT breaks
down on the singular curve where $\varepsilon_{,X}=0$. 

As it follows from (\ref{eq:invertability}), for superluminal speeds
of sound the invertibility of the momentum-velocity relation can be
also violated on a foliation with the relative velocity $v^{2}=1/c_{\text{S}}^{2}$.
In this case the hypersurface of constant time coincides with the
characteristic surface of the equation of motion. Clearly such a surface
cannot be chosen for initial data and the foliation is not suitable
for the Cauchy problem. For a detailed discussion see \cite{Adams:2006sv,Babichev:2007dw}.
Other works discussing the non-uniqueness of the Hamiltonian due to
the multivalued relation between momentum and field velocity include
\cite{Aharonov:1969vu,Henneaux:1987zz,Rizos:2012qs}.

\subsection{Classical and quantum perturbations}

Further the cosmological scalar perturbations are described by the
action 
\begin{equation}
S=\frac{1}{2}\int d\eta d^{3}x\,Z\left(\left(\mathcal{R}'\right)^{2}-c_{\text{S}}^{2}\left(\partial_{i}\mathcal{R}\right)^{2}\right)\,,\label{eq:action for perturbations}
\end{equation}
where $\eta$ is the conformal time and the curvature perturbation
$\mathcal{R}$ is constructed out of the perturbation of the k-essence
field $\delta\varphi$ and the Newtonian potential $\Phi$ 
\begin{equation}
\mathcal{R}=\Phi+H\frac{\delta\varphi}{\dot{\varphi}}\,,\label{eq:curvature_perturbation}
\end{equation}
where both perturbations are written in terms of gauge-invariant variables
which correspond to the Newtonian gage. Finally the normalization
is 
\begin{equation}
Z=a^{2}\left(\frac{\varepsilon+p}{c_{\text{S}}^{2}H^{2}}\right)=\varepsilon_{,X}\left(\frac{\dot{\varphi}a}{H}\right)^{2}\,.\label{eq:Normalization}
\end{equation}
The action is ghosty provided $Z<0$ which is equivalent to $\varepsilon_{,X}<0$.
If NEC is violated, but there are no gradient instabilities so that
the system is hyperbolic and $c_{\text{S}}^{2}>0$, then necessarily
$\varepsilon_{,X}<0$ and the action is ghosty. Ghost instabilities
are perturbative, but rely on interactions with other fields. In particular
there is always an interaction through gravity. On the contrary the
gradient instabilities are linear short scale instabilities. 

If $\varepsilon_{,X}=0$ at the Phantom divide then the background
solution has to go through a singularity of the equation of motion
(\ref{eq:equation_of_motion_cosmology}). For example it is the case
in the example provided in \cite{Nojiri:2006ww}. Even if we assume
that $c_{\text{S}}$ remains nonzero and does not blow up one would
need to guaranty that $\varepsilon_{,\varphi}/\varepsilon_{,X}$ does
not blow up. However, in that case $\varepsilon_{,\varphi}$ should
vanish at the same point where $\varepsilon_{,X}$ is vanishing. This
provides two equations in the phase space $\left(\varphi,\dot{\varphi}\right)$
to satisfy. Therefore this can generically only happen on isolated
points of the measure zero corresponding to extrema of the energy
density. This degenerate case was in details discussed in \cite{Vikman:2004dc}.
Another crucial point is that at the Phantom divide with non-vanishing
speed of sound the normalization factor $Z$ is vanishing. But $Z=0$
corresponds to an infinitely strong coupling on all scales at for
the QFT of perturbations. Indeed, the cubic terms in the action for
perturbations contain higher order derivatives of the background quantities
and are not vanishing at $Z=0$. A QFT which only contains third order
operators is strongly coupled on all scales. 

Another way to understand the pathology is to introduce the canonical
variable (Mukhanov-Sasaki variable) which for for k-essence is \cite{Garriga:1999vw,Mukhanov:2005sc}
\begin{equation}
v=z\mathcal{R}\,,\label{eq:canonical}
\end{equation}
with 
\[
z=\frac{a\dot{\varphi}}{H}\left|\frac{\varepsilon+p}{2Xc_{\text{S}}^{2}}\right|^{1/2}=\frac{a\dot{\varphi}}{H}\,\sqrt{\left|\varepsilon_{,X}\right|}\,.
\]
The canonical variable can be written as 
\begin{equation}
v=a\sqrt{\left|\varepsilon_{,X}\right|}\left(\delta\varphi+\Phi\frac{\dot{\varphi}}{H}\right)\,.\label{eq:canonical_longer}
\end{equation}
The dynamics of this variable is described by the action 

\begin{equation}
S=\frac{1}{2}\sigma\int d\eta d^{3}x\left(\left(v'\right)^{2}+c_{\text{S}}^{2}v\Delta v+\frac{z''}{z}v^{2}\right)\,,\label{eq:action_Canonical}
\end{equation}
$\sigma=\mbox{sign}\left(\varepsilon_{,X}\right)$. The corresponding
dispersion relation is 
\[
\omega_{k}^{2}=c_{\text{S}}^{2}k^{2}-\frac{z''}{z}\,,
\]
where $-z''/z$ plays the role of the square of effective time-dependent
mass. Positive $z''/z$ indicates the Jeans instability operating
on large scales where one cannot neglect expansion of spacetime. This
mass is infinitely growing (in the positive or negative direction)
at the Phantom divide with $\varepsilon_{,X}=0$. Clearly this does
not allow to separate scales and treat the perturbations as an EFT. 

Another problem is rooted in the speed of sound. Let's now assume
that $\varepsilon_{,X}>0$ at the Phantom divide. Then we have $c_{\text{S}}^{2}\left(t_{pd}\right)=0$
and by continuity $c_{\text{S}}^{2}$ changes sign at this point.
This change of sign of $c_{\text{S}}^{2}$ implies the transition
of the equation of motion (\ref{eq:covariant_equation_of_motion})
from the hyperbolic to the elliptic type of PDE or vice versa. It
is well known that the Cauchy problem is ill-posed for elliptical
PDE's. The problem is in the short wavelength instabilities behaving
as 
\begin{equation}
\delta\varphi_{\mathbf{k}}\sim\exp\left(\left|c_{\text{S}}\mathbf{k}\right|t\right)\,,\label{eq:grrowth}
\end{equation}
so that the increment of instability in mode $k$ is $\left|c_{\text{S}}\mathbf{k}\right|$
and grows for large wave numbers without any bound. In reality the
bound is provided by a lattice size or by the strong coupling scale.
In both these cases the instability has a characteristic scale corresponding
to the cut off scale completely destroying the predictability of the
theory. If a system enters the elliptic regime on a classical solution
which is in the formal region of validity of the EFT, than the system
stays there much longer than the inverse frequency cut off scale.
In that case all physical modes get a tremendous exponential amplification
due to (\ref{eq:grrowth}). 

It is worth looking at the quantum fluctuations to understand the
problem. If we start from a healthy background $c_{\text{S}}^{2}>0$
and $\varepsilon_{,X}>0$, then in a Hadamard state (short scales
vacuum state) the mode functions $v_{k}$ for the canonical variable
(\ref{eq:canonical_longer}) are normalized inside the horizon (for
scales $c_{\text{S}}k\ll\left|z''/z\right|$) as 
\begin{equation}
\left|v_{k}\right|=\sqrt{\frac{\hbar}{\omega_{k}}}\simeq\sqrt{\frac{\hbar}{c_{\text{S}}k}}\,.\label{eq:mode_func_norm}
\end{equation}
Where we have explicitly written $\hbar$ to stress the quantum origin
of this quantity. Then the characteristic quantum fluctuation of the
canonical variable on scale $k$ is 
\begin{equation}
\delta v_{k}\sim\sqrt{\hbar}\left|v_{k}\right|k^{3/2}\sim\sqrt{\frac{\hbar}{c_{\text{S}}}}\,k\,.\label{eq:fluctuations}
\end{equation}
On these ultrashort physical length scales $\ell=a/k$ we can neglect
the fluctuations of the Newtonian potential (see below) so that 
\begin{equation}
\delta\varphi_{\ell}\sim\sqrt{\frac{\hbar}{\varepsilon_{,X}c_{\text{S}}}}\left(\frac{k}{a}\right)=\left(\frac{\hbar^{2}}{\varepsilon_{,X}p_{,X}}\right)^{1/4}\cdot\frac{1}{\ell}=\left(\frac{2X\hbar^{2}}{\varepsilon_{,X}\left(\varepsilon+p\right)}\right)^{1/4}\cdot\frac{1}{\ell}\,.\label{eq:field_quantum_fluctuation}
\end{equation}
Here we assumed that $\ell\gg L_{UV}$, where the latter is the UV-cutoff
length scale which is typically present in such derivatively coupled
theories. Thus if either of the quantities $\varepsilon_{,X}$ or
$p_{,X}$ is vanishing, the quantum fluctuations $\delta\varphi_{\ell}$
blow up on short scales. In particular, for $c_{\text{S}}\rightarrow0$
quantum fluctuations at every given scale $\ell$ inside the horizon
blow up and completely invalidate the applicability of the whole theory
of cosmological perturbations. 

On the other hand the fluctuations of the Newtonian potential on short
scales are scale-independent 
\[
\Phi_{\ell}\sim\sqrt{\hbar}\left(\frac{\varepsilon+p}{c_{\text{S}}}\right)^{1/2}\sim\sqrt{\hbar}\left(X\varepsilon_{,X}\left(\varepsilon+p\right)\right)^{1/4}\,,
\]
see page 345 \cite{Mukhanov:2005sc}. These fluctuations are always
small provided $c_{\text{S}}\gtrsim\varepsilon+p$. The latter condition
can also be written in standard units as 
\[
c_{\text{S}}\gtrsim\left(\frac{\varepsilon}{\varepsilon_{\text{Pl}}}\right)\left(1+w\right)\,,
\]
where $\varepsilon_{\text{Pl}}$ is the Planckian energy density\footnote{Clearly this is is a rather weak lower bound on the sound speed for
a dust-like k-essence.}. It is worthwhile to compare the right hand side of this expression
with the Ricci curvature $R=-\varepsilon\left(1-3w\right)$. 

The magnitude of the quantum fluctuations $\delta\varphi_{\ell}$
on a short scale $\ell$ can be also obtained from the following uncertainty
relation 
\begin{equation}
\delta\varphi_{\ell}\cdot\delta P_{\ell}\gtrsim\hbar\,\ell^{-3}\,,\label{eq:uncirtainty}
\end{equation}
where $\delta P_{\ell}$ is the fluctuation of the canonical momentum
on this scale, for a detailed discussion see \cite{Vikman:2012bx}.
Further, the canonical momentum of fluctuations is 
\[
\delta P_{\ell}=G^{tt}\delta\dot{\varphi}_{\ell}=\varepsilon_{,X}\delta\dot{\varphi}_{\ell}\,.
\]
The fluctuation of the field velocity on short-scale $\ell$ can be
estimated as 
\[
\delta\dot{\varphi}_{\ell}\simeq\omega_{\ell}\delta\varphi_{\ell}\simeq\left(c_{\text{S}}/\ell\right)\delta\varphi_{\ell}\,.
\]
For an oscillator the vacuum saturates the uncertainty relation therefore
for a collection of oscillators 
\[
\delta\varphi_{\ell}^{2}\simeq\frac{\hbar}{\varepsilon_{,X}c_{\text{S}}}\cdot\frac{1}{\ell^{2}}\,,
\]
which again gives (\ref{eq:field_quantum_fluctuation}). This estimation
is not applicable, if the system is strongly coupled for scales $c_{\text{S}}k\lesssim\left|z''/z\right|$.
However, in that case the predictive power of such theory is rather
limited.

\section{General phase space analysis and the Bendixson\textendash Dulac theorem\label{sec:General-phase-space}}

The cosmological equation of motion (\ref{eq:equation_of_motion_cosmology})
can be written in the first order form 
\begin{align}
 & \frac{d\dot{\varphi}}{dt}=-3H\dot{\varphi}c_{\text{S}}^{2}-\frac{\varepsilon_{,\varphi}}{\varepsilon_{,X}}\,,\label{eq:first_order_system_norm}\\
 & \frac{d\varphi}{dt}=\dot{\varphi}\,.\nonumber 
\end{align}
The first equation of this system above is singular for configurations
with $\varepsilon_{,X}=0$. The integral curves for the vector field
$\left(\dot{\varphi},\ddot{\varphi}\right)$ of on the phase space
$\left(\varphi,\dot{\varphi}\right)$ are given by 
\begin{equation}
\frac{d\dot{\varphi}}{d\varphi}=-3Hc_{\text{S}}^{2}-\frac{\varepsilon_{,\varphi}}{\varepsilon_{,\dot{\varphi}}}\,.\label{eq:integral_curves}
\end{equation}
For every k-essence with the Lagrangian $p\left(\varphi,\dot{\varphi}\right)$
it is convenient to introduce an auxiliary system of ODE \cite{Vikman:2004dc}
\begin{align}
 & \frac{du}{dt}=\alpha\left(u,v\right)=-p_{,u}\sqrt{3\varepsilon}-\varepsilon_{,v}\,,\label{eq:auxiliarly_system}\\
 & \frac{dv}{dt}=\beta\left(u,v\right)=\varepsilon_{,u}\,,\nonumber 
\end{align}
 with $\varepsilon=up_{,u}-p$. Without the friction term $p_{,u}\sqrt{3\varepsilon}$
these ODEs are the Hamilton equations of motion with canonical momentum
$u$, coordinate $v$ and Hamiltonian $\varepsilon$. Without this
friction term the motion happens on the curves of constant $\varepsilon$.
As $p_{,u}=up_{,X}$, the effective friction coefficient is $p_{,X}\sqrt{3\varepsilon}$.
If the sign of the friction coefficient is always positive the dissipative
motion cannot be periodic. Hence $p_{,X}$ has to change the sign
for the existence of a limiting cycle. This is a reformulation of
the same statement presented in at the beginning of the paper. This
system (\ref{eq:auxiliarly_system}) is different from (\ref{eq:equation_of_motion_cosmology})
but is constructed out of the same function $p\left(\varphi,\dot{\varphi}\right)$
where instead of $\varphi$ one plugs in $v$ and instead of $\dot{\varphi}$
one plugs in $u$. The solutions $u\left(t\right)$ and $v\left(t\right)$
are different from $\varphi\left(t\right)$ and $\dot{\varphi}\left(t\right)$.
In particular, $\dot{v}\neq u$. The main point is that this system
(\ref{eq:auxiliarly_system}) has the same integral curves on $\left(v,u\right)$
as the equation of motion (\ref{eq:equation_of_motion_cosmology})
on $\left(\varphi,\dot{\varphi}\right)$, but it is clearly less singular.
This integral curves are locally given by (\ref{eq:integral_curves}).
In particular a limiting cycle corresponds to a closed integral curve
on $\left(\varphi,\dot{\varphi}\right)$ plane and on $\left(v,u\right)$
plane. The time flow in $\left(v,u\right)$ goes
in the opposite direction to the time flow in $\left(\varphi,\dot{\varphi}\right)$
for the regions with $\varepsilon_{,X}<0$ as both first order equations
(\ref{eq:auxiliarly_system}) for $v$ and $u$ have an opposite sign
to the corresponding equations (\ref{eq:first_order_system_norm}). 

Then by the Bendixson\textendash Dulac theorem, if there exist a $C^{1}$
function $f\left(u,v\right)$ (called the Dulac function) such that
\[
\frac{\partial\left(f\alpha\right)}{\partial u}+\frac{\partial\left(f\beta\right)}{\partial v}>0\,,
\]
almost everywhere in a simply connected region of the plane, then
there are no periodic solutions lying entirely within the region. 

Further we will assume that in the region where of phase space where
the limiting cycle is located the energy density is strictly positive,
$\varepsilon>0$. This is needed for the differentiability of the
right hand side of the system (\ref{eq:auxiliarly_system}). Moreover,
this excludes a possibility of a bounce happening in this region of
phase space. 

In that case one can chose the Dulac function $f\left(\varepsilon\right)=1/\sqrt{3\varepsilon}$
for which we have\footnote{It is important to stress the power of the Dulac generalization of
the original Bendixson theorem which only allowed for $f=1$. In that
case one would obtain that $\varepsilon_{,X}\left(3\varepsilon+p\right)$
should change the sign for the existence of the limiting cycle. This
is clearly a weaker requirement. } 
\[
\frac{\partial\left(f\alpha\right)}{\partial u}+\frac{\partial\left(f\beta\right)}{\partial v}=\left(-f\left(\varepsilon\right)p_{,u}\sqrt{3\varepsilon}\right)_{,u}=-\varepsilon_{,X}\,,
\]
where we have used that $p_{,uu}=\left(up_{,X}\right)_{,u}=p_{,X}+2Xp_{,XX}=\varepsilon_{,X}$.
Hence the system should necessary possess a singularity where $\varepsilon_{,X}$
changes sign or at least vanishing. Otherwise the limiting cycle cannot
exist. This goes beyond the statement that the limiting cycle should
have a fixed point of the system (\ref{eq:auxiliarly_system}) inside.
Indeed, for a fixed point $\varepsilon_{,u}=\varepsilon_{,X}u$ should
vanish. The latter is always realized at the origin where $u=0$.
Note that the change of sign of $\varepsilon_{,X}$ is a strong requirement
for the existence of a limiting cycle. It is not clear whether this
condition can be obtained just form the requirement that the sign
of the time-derivative of the pressure (\ref{eq:derivative_of_pressure})
should change. 

It is important to note that $u$ does not correspond to a naive \emph{flat
space} canonical momentum $\pi=p_{,\dot{\varphi}}=P/a^{3}$
for the field $\varphi$. Indeed, for this momentum $\pi$ we obtain
using (\ref{eq:equation_of_motion_cosmology}) 
\begin{equation}
\dot{\pi}=-3H\pi+\left(\frac{\partial p}{\partial\varphi}\right)_{\dot{\varphi}}\,,\label{eq:fake_canonical}
\end{equation}
 instead of the first equation of the auxiliary system (\ref{eq:auxiliarly_system}).
From this equation above it follows that on the limiting cycle the
$p_{,\varphi}$ (taken by constant $\dot{\varphi}$) should change
the sign. Thus a limiting cycle crosses the curve on phase space $\left(\varphi,\dot{\varphi}\right)$
where $p_{,\varphi}=0$. Indeed, otherwise it is impossible to overcome
the Hubble friction and the\emph{ flat space} canonical momentum $\pi$
would always decrease in an expanding universe. Note that the transition
to $\pi$ is not that useful in the region with a singularity $\varepsilon_{,X}=0$,
as there the relation between momentum and field velocity becomes
not invertible, see (\ref{eq:invertability}). Thus this description
would not be useful to prove the statement that this singularity is
required for a limiting cycle. Now we can look at the \emph{flat space}
Hamiltonian 
\begin{equation}
\mathcal{H}=\dot{\varphi}\pi-p\,.\label{eq:another_Hamiltonian}
\end{equation}
Clearly this Hamiltonian (\ref{eq:another_Hamiltonian}) is equal
to $\varepsilon$ by value, but $\varepsilon\left(u,v\right)$ and
$\mathcal{H}\left(\varphi,\pi\right)$ are two different functions.
Further using 
\[
\left(\frac{\partial p}{\partial\varphi}\right)_{\pi}=\left(\frac{\partial p}{\partial\varphi}\right)_{\dot{\varphi}}+\left(\frac{\partial p}{\partial\dot{\varphi}}\right)_{\varphi}\left(\frac{\partial\dot{\varphi}}{\partial\varphi}\right)_{\pi}\,,
\]
we get that 
\[
\left(\frac{\partial p}{\partial\varphi}\right)_{\dot{\varphi}}=-\left(\frac{\partial\mathcal{H}}{\partial\varphi}\right)_{\pi}\,,
\]
so that (\ref{eq:fake_canonical}) differs from the canonical equation
by a friction term with\emph{ }$3H=\sqrt{3\mathcal{H}}$ friction
coefficient. The first order equation for $\varphi$ does not have
the friction term and remains canonical 
\[
\dot{\varphi}=\left(\frac{\partial\mathcal{H}}{\partial\pi}\right)_{\varphi}\,.
\]
Contrary to the auxiliary system (\ref{eq:auxiliarly_system}) the
trajectories in $\left(\varphi,\pi\right)$ are not identical to the
integral curves of (\ref{eq:equation_of_motion_cosmology}). 

We can start from the system 
\begin{align}
 & \dot{\pi}=-\sqrt{3\mathcal{H}}\,\pi-\mathcal{H}_{,\varphi}\,,\label{eq:almoust_Hamilton_equations}\\
 & \dot{\varphi}=\mathcal{H}_{,\pi}\,,\nonumber 
\end{align}
and use the Bendixson\textendash Dulac theorem with the Dulac function
$f=1/\sqrt{3\mathcal{H}}$ to obtain that $\partial_{\pi}\left(f\alpha\right)+\partial_{\varphi}\left(f\beta\right)=-1$.
Which implies that the limiting cycle is not possible, provided the
conditions of the Bendixson\textendash Dulac theorem are satisfied.
In particular, the relevant assumption there was that $\alpha\left(\varphi,\pi\right)$
and $\beta\left(\varphi,\pi\right)$ are smooth - i.e. that the flat
space Hamiltonian $\mathcal{H}$ is smooth. Hence a limiting cycle
is not possible for k-essence with a smooth Hamiltonian. Moreover,
one can expect that the \emph{cosmocanonical} system (\ref{eq:almoust_Hamilton_equations})
is universal \textendash{} holds for \emph{any generic} cosmological
scalar field \emph{beyond} k-essence i.e. Galileons, Horndeski theories
etc \footnote{We assume here that the system (\ref{eq:almoust_Hamilton_equations})
is valid for a cosmological scalar field with a second order covariant
equation of motion. A detailed discussion will be provided somewhere
else. }. Therefore we expect that the limiting cycle in an expanding universe
is only possible for systems with a not smooth flat space Hamiltonian.

\section{Phase space analysis of the particular model\label{sec:Particular_Model}}

Here we provide a phase space analysis of the class of k-essence systems
discussed in \cite{Bains:2015gpv}. The expansion of the universe
will be assumed to be driven by the scalar field itself through the
first Friedmann equation (\ref{eq:Friedmann_1}). This analysis will
mostly serve for the illustrative purposes of our generic statements
made above. The Lagrangian studied in \cite{Bains:2015gpv} is 

\begin{equation}
p(\varphi,X)=\left(3b\,\varphi^{2}-1\right)X+X^{2}-V\left(\varphi\right)\,,\label{eq:kosc}
\end{equation}
where the double-well potential is given by 
\begin{equation}
V\left(\varphi\right)=\Lambda+\frac{1}{12a}-\frac{1}{2}\varphi^{2}+\frac{3a}{4}\varphi^{4}\,.\label{eq:potential}
\end{equation}
Without potential and with $b=0$ this system corresponds to simple
k-inflation \cite{ArmendarizPicon:1999rj} and ghost condensate \cite{ArkaniHamed:2003uy}.
The authors only considered $a>0$ and $b>0$, which we will assume
in this paper as well. The extrema of the potential are given by 
\begin{align}
\varphi_{0}=0\,, & \:\varphi_{\pm}=\pm\frac{1}{\sqrt{3a}}\,.\label{eq:Fixed_Points}
\end{align}
Thus on in the phase space of the homogeneous solutions $\left(\varphi,\dot{\varphi}\right)$
there are always three equilibrium points - fixed points $\left(0,0\right)\,,\left(\pm1/\sqrt{3a},0\right)$.
Clearly the first trivial point $\left(0,0\right)$ without symmetry
breaking is unstable. There $V\left(\varphi_{0}\right)=\Lambda+1/\left(12a\right)$.
While the fixed points with symmetry breaking correspond to de Sitter
solutions with 
\[
V\left(\varphi_{\pm}\right)=\Lambda\,.
\]
The NEC is violated when $p_{,X}<0$ which occurs inside of the phantom
divide which is given by the ellipse 
\begin{equation}
\dot{\varphi}^{2}+3b\,\varphi^{2}=1\,.\label{eq:NEC_ellipse}
\end{equation}
For values of $\varphi$ and $\dot{\varphi}$ outside of the ellipse
the system does not violate the NEC. On the ellipse the equation of
state $w=-1$. From (\ref{eq:field_quantum_fluctuation}) it follows
that by approaching the phantom divide ellipse from outside the quantum
perturbations on short scales diverge. 

The energy density is given by (\ref{eq:Energy_density}) 
\[
\varepsilon=\left(3b\,\varphi^{2}-1\right)X+3X^{2}+V\left(\varphi\right)\,.
\]
The singularity in the equation of motion (\ref{eq:equation_of_motion_cosmology})
occurs when $\varepsilon_{,X}=0$ or on the ellipse 
\begin{equation}
3\dot{\varphi}^{2}+3b\,\varphi^{2}=1\,.\label{eq:Singularity_ellipse}
\end{equation}
From (\ref{eq:field_quantum_fluctuation}) it follows that by approaching
this singularity ellipse with $\varepsilon_{,X}=0$ from inside the
quantum perturbations on short scales diverge. Note that both ellipses
(\ref{eq:NEC_ellipse}) and (\ref{eq:Singularity_ellipse}) share
the same $\varphi$ axis, but the axis in $\dot{\varphi}$ for the
NEC violation is in $\sqrt{3}$ times larger. Hence the singularity
ellipse (\ref{eq:Singularity_ellipse}) is always inside of the NEC-violation
ellipse (\ref{eq:NEC_ellipse}). 

For the sound speed (\ref{eq:sound}) as a function on phase space
we have 
\[
c_{\text{S}}^{2}\left(\varphi,\dot{\varphi}\right)=\frac{\dot{\varphi}^{2}+3b\,\varphi^{2}-1}{3\dot{\varphi}^{2}+3b\,\varphi^{2}-1}=1-\frac{2\dot{\varphi}^{2}}{3\dot{\varphi}^{2}+3b\,\varphi^{2}-1}\,.
\]
Thus for all $\varepsilon_{,X}=3\dot{\varphi}^{2}+3b\,\varphi^{2}-1<0$
i.e. inside of the ellipse (\ref{eq:Singularity_ellipse}) the sound
speed is always superluminal except of points $\dot{\varphi}=0$.
For $\varepsilon_{,X}>0$ the speed of sound is never superluminal.
By approaching this singular ellipse (\ref{eq:Singularity_ellipse})
from inside the sound speed grows without any limit. It is important
that the superluminality is separated from a limiting cycle by the
singularity ellipse (\ref{eq:Singularity_ellipse}). Indeed, no trajectory
can cross this border, while we know that the limiting cycle (if exists)
has to cross the Phantom divide ellipse (\ref{eq:NEC_ellipse}) which
is located outside of the singularity ellipse. In the region of the
phase space between the NEC-violation ellipse (\ref{eq:NEC_ellipse})
and the singularity ellipse (\ref{eq:Singularity_ellipse}) the sound
speed is imaginary and the the system is elliptic. The increment of
this linear instability grows for each mode without any bound by approaching
the singularity ellipse (\ref{eq:Singularity_ellipse}). The linear
instability for high $k$ modes was mentioned in \cite{Bains:2015gpv},
albeit the authors used a different formula for the sound speed. Namely
it was assumed that one can treat the system as a fluid so that the
sound speed can be inferred from $c_{s}^{2}=\partial\left\langle p\right\rangle /\partial\left\langle \varepsilon\right\rangle $
where the averaging is done for many oscillations. Clearly this formula
is different from (\ref{eq:sound}). 

On the other hand for the derivative of the energy density we have
\[
\varepsilon_{,\varphi}=\varphi\left(3b\dot{\varphi}^{2}+3a\varphi^{2}-1\right)\,.
\]
Thus there is a ellipse of the vanishing $\varepsilon_{,\varphi}$
\[
3b\dot{\varphi}^{2}+3a\varphi^{2}=1\,.
\]
As we have showed in the section (\ref{sec:General-phase-space})
the limiting cycle has to go through the curve $p_{,\varphi}=0$.
This curve is given by 
\[
p_{,\varphi}=\varphi\left(3b\dot{\varphi}^{2}+1-3a\varphi^{2}\right)\,.
\]
Thus it is enough to evolve through the $\varphi=0$. 

\begin{figure}[tb]
\begin{centering}
\includegraphics[width=7.5cm]{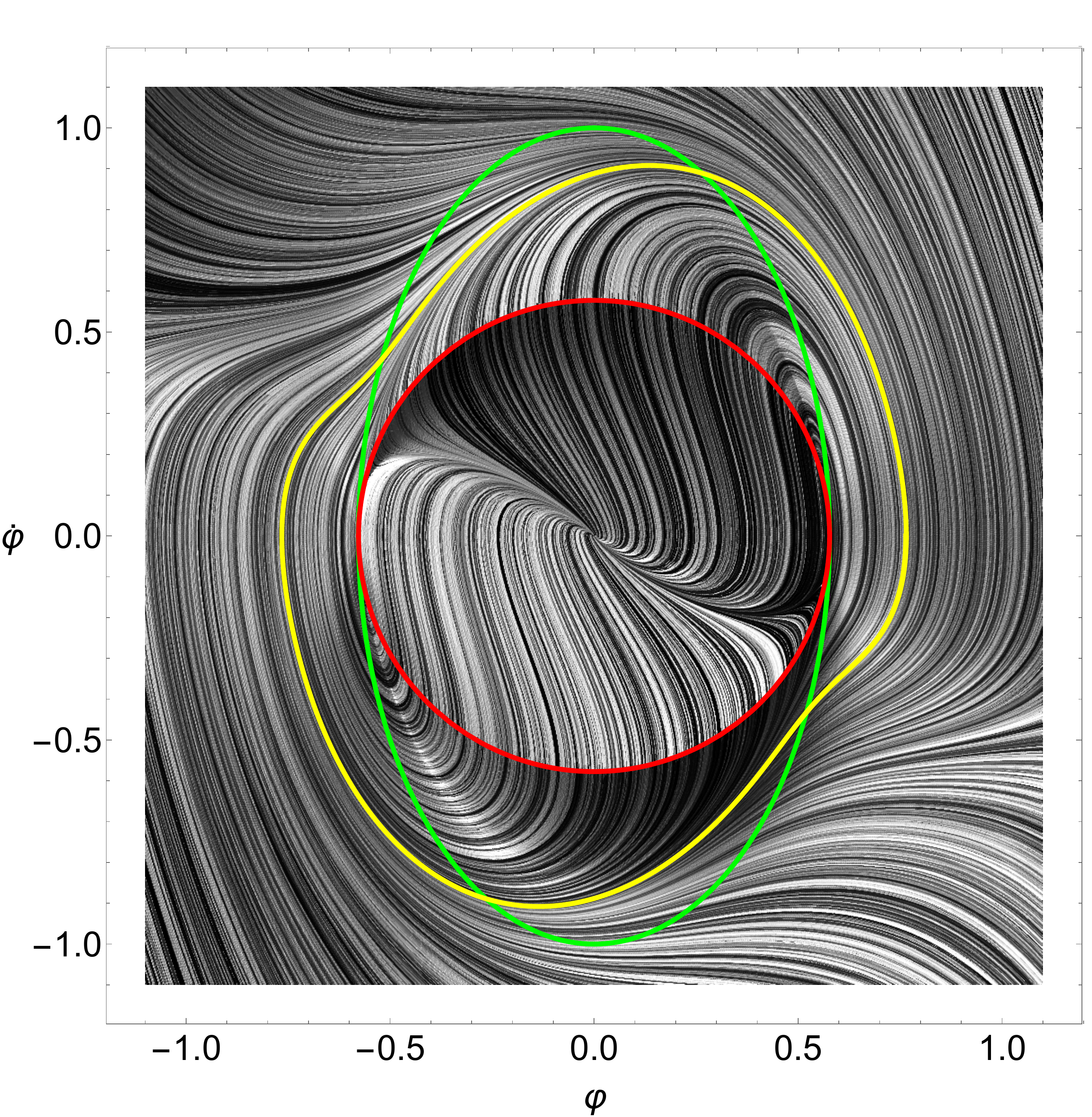}\includegraphics[width=7.5cm]{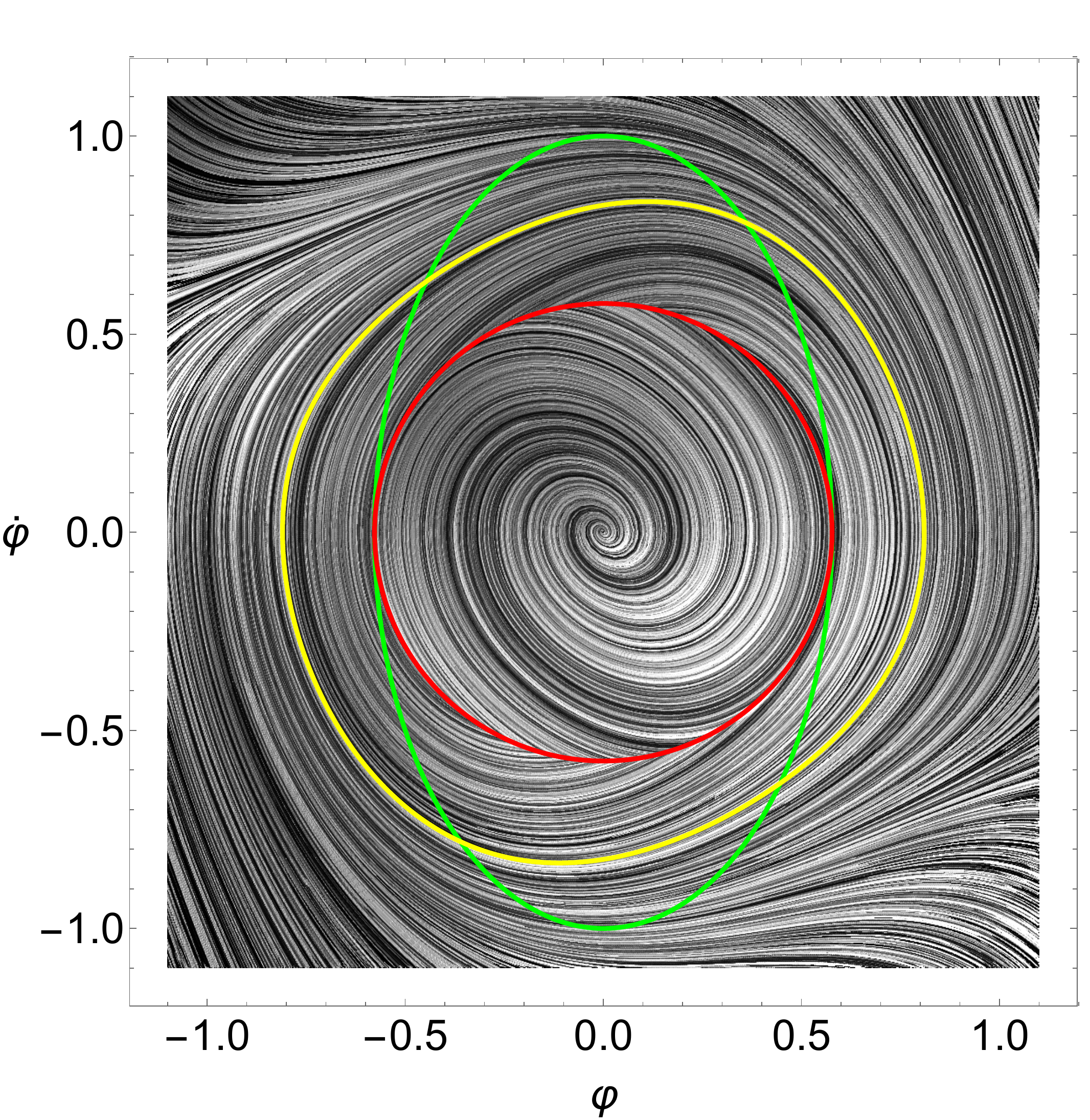}
\par\end{centering}

\caption{\label{fig:Limit_2}\foreignlanguage{american}{ Phase plot with a
limiting cycle for the system (\ref{eq:kosc}) with the degenerate
values of the parameters: $a=1$, $b=1$ and $\Lambda=0$ on the right
and $a=1$, $b=1$ and $\Lambda=1$ on the left. The values of parameters
are chosen for illustrative purposes only. The limiting cycle is the
yellow less-symmetric curve. The red ellipse is the singularity where
$\varepsilon_{,X}=0$, which is given by (\ref{eq:Singularity_ellipse}).
Inside of this singular curve the perturbations are ghosts. On the
singular curve the perturbations are infinitely strongly coupled.
The green ellipse is the curve on which $p_{,X}=0$ and $w=-1$. This
green ellipse is given by (\ref{eq:NEC_ellipse}). Inside this green
ellipse the NEC is violated. Between these two ellipses the sound
speed in imaginary and the system is linearly unstable. By approaching
the red ellipse from outside this linear instability becomes infinitely
strong for any given mode $k$. }\protect \\
\foreignlanguage{american}{Clearly both systems possess a trivial
equilibrium / fix point at the origin $\left(\varphi,\dot{\varphi}\right)=\left(0,0\right)$
where the system is linearly stable but is ghosty and suffers therefore
from the ghosts instabilities due to interactions (e.g. unavoidable
interactions through gravity). This trivial vacuum is isolated by
the red ellipse of the infinitely strong coupling from the oscillatory
attractor which can be considered as a nontrivial ground state. However,
this is nontrivial ground state is plagued by linear instabilities.
Indeed, the yellow cyclic trajectory clearly crosses the green ellipse
four times - enters twice the region where the sound speed is imaginary.
Moreover, the classical evolution breaks down by approaching the green
ellipse as the quantum perturbations on short scales grow without
any bound, see (\ref{eq:field_quantum_fluctuation}). }\protect \\
\foreignlanguage{american}{}\protect \\
\protect \\
\protect \\
}
\end{figure}

For our further analysis of the phase curves it is convenient to use
the auxiliary system(\ref{eq:auxiliarly_system}) 
\begin{align}
 & \frac{du}{dt}=-p_{,u}\sqrt{3\varepsilon}-\varepsilon_{,v}=-u\left(3bv^{2}+u^{2}-1\right)\sqrt{3\varepsilon}-v\left(3bu^{2}+3av^{2}-1\right)\,,\label{eq:auxiliarly_system_concrete_example}\\
 & \frac{dv}{dt}=\varepsilon_{,u}=u\left(3bv^{2}+3u^{2}-1\right)\,,\nonumber 
\end{align}

where the energy density 
\begin{equation}
\varepsilon\left(v,u\right)=\frac{1}{2}\left(3b\,v^{2}-1\right)u^{2}+\frac{3}{4}u^{4}+\Lambda+\frac{1}{12a}-\frac{1}{2}v^{2}+\frac{3a}{4}v^{4}\,.\label{eq:energy_density_uv}
\end{equation}
For this system the singular ellipse (\ref{eq:Singularity_ellipse})
does not correspond to any singularity. In generic case when $a\neq b\neq1$
the flow $\overrightarrow{F}=\left(\dot{v},\dot{u}\right)$ on the
singular ellipse is parallel to $u$-axis, see Fig. (\ref{fig:Limit_2})
for a degenerate case. For small $v$ or close to the $u$-axis the
flow is always pointing out outside of the singular ellipse, as the
cosmological friction dominates. However, by approaching the $v$-axis
the second term in the equation (\ref{eq:auxiliarly_system_concrete_example})
starts to dominate. And the direction of flow changes inwards. The
point of equilibrium is between these two forces is a nontrivial fixed
point. 

\begin{figure}[tb]
\begin{centering}
\includegraphics[width=14.5cm]{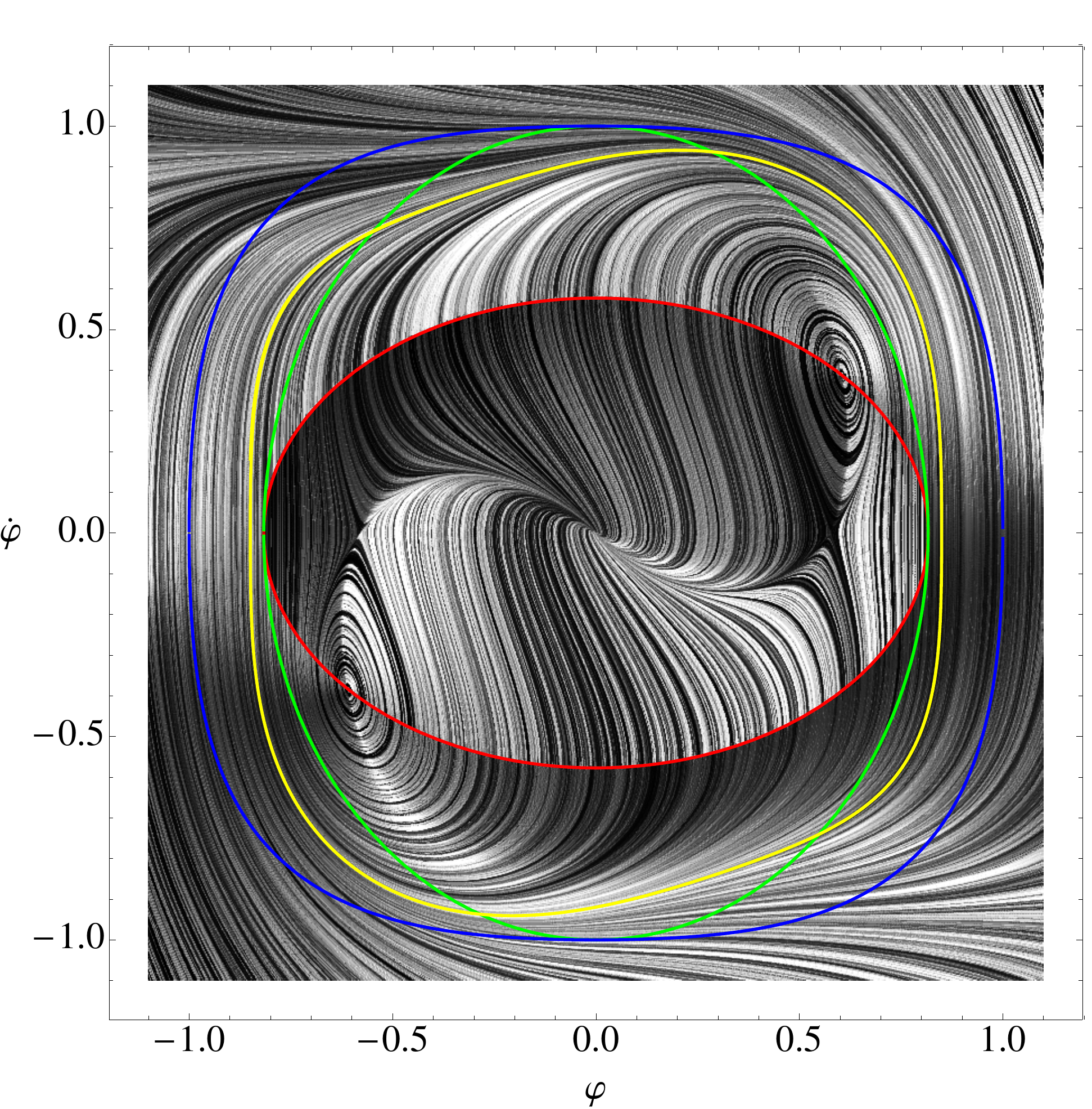}
\par\end{centering}

\caption{\label{fig:Limit_1}{ Phase plot with a
limiting cycle for the system (\ref{eq:kosc}) with the parameters
$a=1$, $b=1/2$ and $\Lambda=1$ The values of parameters are chosen
for illustrative purposes only. The limiting cycle is the yellow less-symmetric
curve. The red ellipse is the singularity where $\varepsilon_{,X}=0$,
which is given by (\ref{eq:Singularity_ellipse}). Inside of this
singular curve the perturbations are ghosts. On the singular curve
the perturbations are infinitely strongly coupled. The green ellipse
is the Phantom divide on which $p_{,X}=0$ and $w=-1$. This green
ellipse is given by (\ref{eq:NEC_ellipse}). Inside this green ellipse
the NEC is violated. Between these two ellipses the sound speed in
imaginary and the system is linearly unstable. By approaching the
red ellipse from outside this linear instability becomes infinitely
strong for any given mode $k$. The blue contour is the lowest level
of constant energy density enclosing the Phantom divide, given by
(\ref{eq:focusing_contour}). One can clearly see five fixed points:
two focuses on the singular ellipse, one node in the origin and two
saddles on the $\varphi$-axis. }\protect \\
\foreignlanguage{american}{Clearly the system possesses a trivial
equilibrium / fix point at the origin $\left(\varphi,\dot{\varphi}\right)=\left(0,0\right)$
where the system is linearly stable but is ghosty and suffers therefore
from the ghosts instabilities due to interactions (e.g. unavoidable
interactions through gravity). This trivial vacuum is isolated by
the red ellipse of the infinitely strong coupling from the yellow
oscillatory attractor which can be considered as a nontrivial ground
state. However, this nontrivial ground state is plagued by linear
instabilities. Indeed, the yellow cyclic trajectory clearly crosses
the green ellipse four times - enters twice the region where the sound
speed is imaginary. Moreover, the classical evolution breaks down
by approaching the green ellipse as the quantum perturbations on short
scales grow without any bound, see (\ref{eq:field_quantum_fluctuation}).
}\protect \\
}
\end{figure}

Without the friction term the motion occurs in $\left(v,u\right)$
along the contours of the constant energy density. If we find an energy
level completely enclosing the phantom divide from the outside where
$p_{,X}>0$ (so that in this model $\varepsilon_{,X}>0$ too) the
flow of the time evolution in an expanding universe will be directed
to the interior phase space of this level of energy density. Indeed,
the normal to the $\varepsilon\left(v,u\right)=\varepsilon_{0}$ contour
is $\overrightarrow{N}=\left(\varepsilon_{,v},\varepsilon_{,u}\right)$,
it points out outside of the closed contour, while the scalar product
of a $\overrightarrow{N}$ with flow $\overrightarrow{F}=\left(\dot{v},\dot{u}\right)$
is 
\[
\overrightarrow{F}\cdot\overrightarrow{N}=-\varepsilon_{,u}p_{,u}\sqrt{3\varepsilon}=-2X\varepsilon_{,X}p_{,X}\sqrt{3\varepsilon}\,.
\]
It is easy to check that the curve of constant energy density 
\begin{equation}
\varepsilon\left(v,u\right)=\varepsilon_{pd}=\frac{1}{4}+\frac{1}{12a}+\Lambda\,,\label{eq:focusing_contour}
\end{equation}
where $\varepsilon\left(v,u\right)$ is given by (\ref{eq:energy_density_uv})
has the smallest energy density among the levels encircling the Phantom
divide ellipse (\ref{eq:NEC_ellipse}), see Fig. (\ref{fig:Limit_1}).
Thus any infinitesimally small change of the contour $\varepsilon\left(v,u\right)=\varepsilon_{pd}\left(1+\epsilon\right)$
where $\epsilon\ll1$ encloses the Phantom divide and has the flow
of trajectories pointing inside. No trajectory can escape this contour
(\ref{eq:focusing_contour}). As a limiting cycle has to cross the
Phantom divide, there cannot be limiting cycles outside of the contour
of no return. This contour of no return crosses the $v$-axes at 
\begin{equation}
\varphi_{nr\pm}=\pm\sqrt{\frac{1+\sqrt{1+3a}}{3a}}\,.\label{eq:Phi_nr}
\end{equation}

The limiting cycle cannot cross the singularity ellipse (\ref{eq:Singularity_ellipse}).
Hence we have the following bound on the amplitude of the oscillations
of the field 
\[
\frac{1}{\sqrt{3b}}<\varphi_{max}<\sqrt{\frac{1+\sqrt{1+3a}}{3a}}\,.
\]
Let us study the fixed points in details. We start from the trivial
fixed point $\left(u,v\right)=0$ in that case 
\[
\varepsilon=\Lambda+\frac{1}{12a}+\mathcal{O}\left(u^{2},v^{2}\right)=V_{0}+\mathcal{O}\left(u^{2},v^{2}\right)\,,
\]
so that the linearized system takes the form 
\begin{align}
 & \frac{du}{dt}=u\sqrt{3V_{0}}+v\,,\label{eq:Origin_Fixed}\\
 & \frac{dv}{dt}=-u\,,\nonumber 
\end{align}
with the corresponding eigenvalues 
\[
\lambda_{\pm}^{0}=\sqrt{\frac{3V_{0}}{4}}\pm\sqrt{\frac{3V_{0}}{4}-1}\,.
\]
For $V_{0}\geq4/3$ both eigenvalues are real and positive so that
the fixed point is an unstable node. While for $V_{0}<4/3$ both root
are complex with a positive real part so that the fixed point is an
unstable focus. In both cases all trajectories leave the small neighborhood
of the trivial fixed point. 

\begin{figure}[tb]
\begin{centering}
\includegraphics[width=14.5cm]{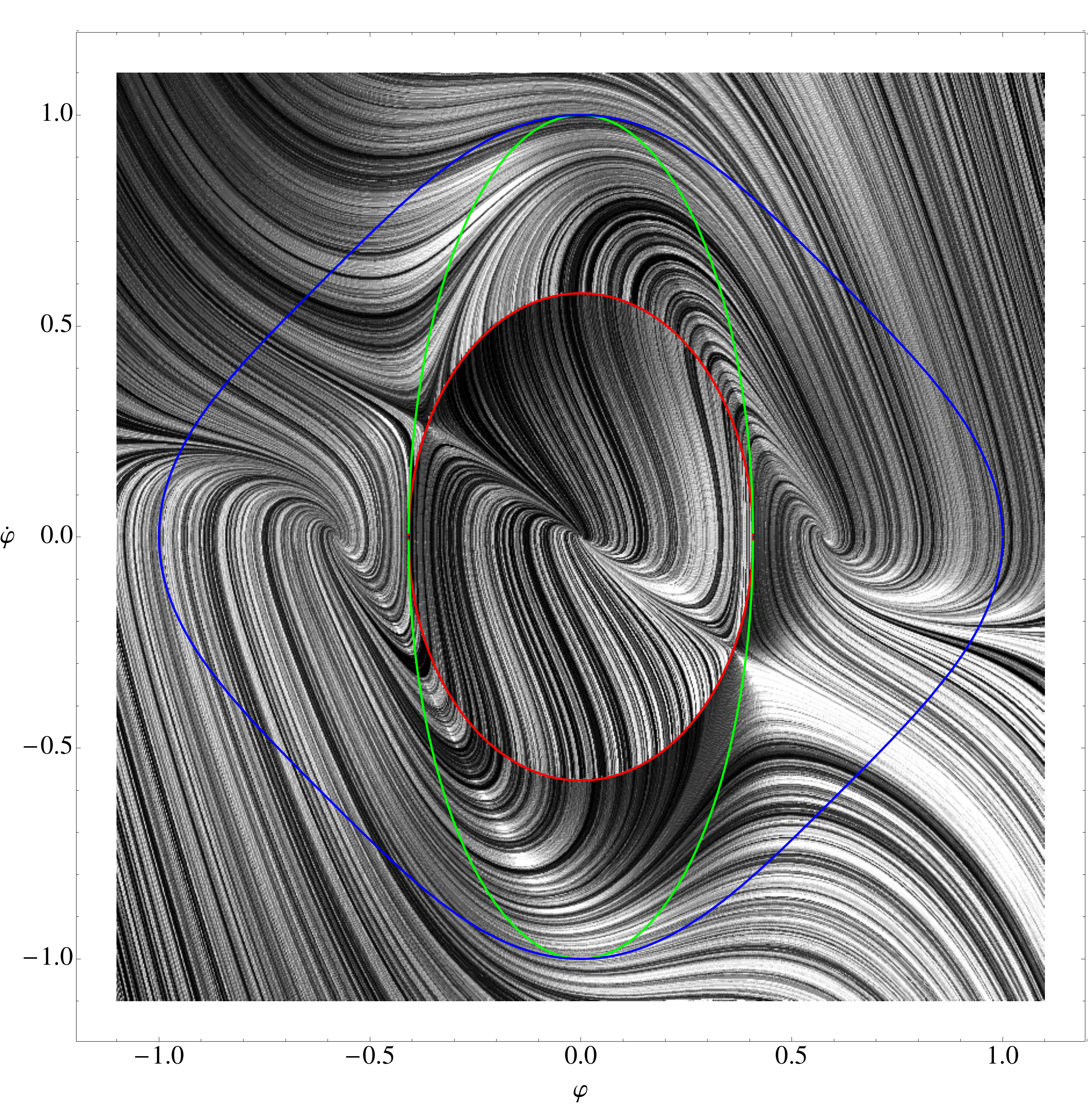}
\par\end{centering}

\caption{\label{fig:No_Limit_3}\foreignlanguage{american}{ Phase plot without
the limiting cycle for the system (\ref{eq:kosc}) with the values
of the parameters $a=1$, $b=2$ and $\Lambda=1$ on the left. The
values of parameters are chosen for illustrative purposes only. The
red ellipse is the singularity where $\varepsilon_{,X}=0$, which
is given by (\ref{eq:Singularity_ellipse}). Inside of this singular
curve the perturbations are ghosts. On the singular curve the perturbations
are infinitely strongly coupled. The green ellipse is the curve on
which $p_{,X}=0$ and $w=-1$. This green ellipse is given by (\ref{eq:NEC_ellipse}).
Inside this green ellipse the NEC is violated. Between these two ellipses
the sound speed in imaginary and the system is linearly unstable.
By approaching the red ellipse from outside this linear instability
becomes infinitely strong for any given mode $k$. The blue contour
is the lowest level of constant energy density enclosing the Phantom
divide, given by (\ref{eq:focusing_contour}). One can clearly see
five fixed points: two saddles on the singular ellipse, one node in
the origin and two stable fixed points on the $\varphi$-axis. These
fixed points are stable focuses and attract trajectories crossing
the blue contour of no return. Outside of the blue contour there cannot
be limiting cycles as the latter has to cross through the green ellipse
- the Phantom divide. }\protect \\
\protect \\
}
\end{figure}

Now let's consider other fixed points $\left(0,\pm\left(3a\right)^{-1/2}\right)$.
Because of (\ref{eq:Phi_nr}) these fixed points are always inside
of the no return contour (\ref{eq:focusing_contour}). If these points
are stable and are outside of the Phantom divide $b>a$, they allow
to trajectories to end up without creating the limiting cycle. There
we have for the energy 
\[
\varepsilon=\Lambda+\mathcal{O}\left(\delta u^{2},\delta v^{2}\right)\,,
\]
so that the linearized system is 
\begin{align}
 & \frac{du}{dt}=-u\left(\frac{b}{a}-1\right)\sqrt{3\Lambda}-2\delta v\,,\label{eq:nontrivial_fixed_point}\\
 & \frac{d\delta v}{dt}=u\left(\frac{b}{a}-1\right)\,.\nonumber 
\end{align}
The corresponding eigenvalues are 
\[
\lambda_{\pm}^{c}=\frac{1}{2}\left[\left(1-\frac{b}{a}\right)\sqrt{3\Lambda}\pm\sqrt{3\Lambda\left(1-\frac{b}{a}\right)^{2}+8\left(1-\frac{b}{a}\right)}\right]\,.
\]
If $b>a$ both roots have negative real part. In that case $\varphi_{\pm}$
is outside of the phantom divide and both fixed points are stable
focuses, see (\ref{fig:No_Limit_3}). In that case trajectories starting
crossing the no return contour (\ref{eq:focusing_contour}) from outside
can end on these stable focuses. If $b<a$ then both roots are real
and $\lambda_{+}^{c}>0$ while $\lambda_{-}^{c}<0$. Hence the fixed
points are saddle points i.e unstable for $b<a$. Thus for a limiting
cycle we need that $b<a$. 

Other nontrivial fixed points are located at the singular ellipse.

\selectlanguage{american}%

\section{Conclusions and Discussion }

The new oscillatory state of cosmological matter corresponds to a
limiting cycle of the classical equations of motion. This state can
also be considered as a cosmological realization of a time-crystal.
This regime can be of interest to cosmology, because on average the
universe undergoes the de Sitter expansion with $\left\langle w\right\rangle =-1$.
Hence, a time-crystal can be used to model early stages of Inflation
or late stages of Dark Energy. We have shown that any realization
of the new oscillatory state of cosmological matter requires not only
a violation of the NEC but also a crossing of the Phantom divide.
The systems studied in \cite{Bains:2015gpv} are known to be incapable
to achieve this crossing \cite{Vikman:2004dc}, see also \cite{Hu:2004kh,Caldwell:2005ai,Xia:2007km}
and for a review \cite{Cai:2009zp,Nesseris:2006er}. In particular,
the crossing generically implies gradient instabilities where the
sound speed is imaginary, $c_{s}^{2}<0$. We have showed in section
\ref{sec:Refreshing-k-essence-and} that the quantum perturbations
of k-essence on short scales grow without any bound in an attempt
to evolve across $w=-1$. 

Further in section (\ref{sec:General-phase-space}) we used the Bendixson\textendash Dulac
theorem to prove that for k-essence i) to realize a cosmological limiting
cycle classically the system has to have a flat-space Hamiltonian
with cusps and that ii) the existence of a limiting cycle implies
that the system has a singularity where $\varepsilon_{,X}$ changes
sign and where the canonical momentum does not define the velocity
uniquely. The appearance of this singularity implies the presence
of configurations with strong superluminal propagation of the small
perturbations and existence of other configurations with strong gradient
instability. On this singularity the small quantum perturbations are
infinitely strongly coupled. The Bendixson\textendash Dulac theorem
dictates that the singularity curve on which $\varepsilon_{,X}$ is
vanishing, should be located inside of the limiting cycle, as the
latter cannot cross the singularity. For some systems, including the
one studied in \cite{Bains:2015gpv}, the configurations with superluminality
are separated from the limiting cycle by the singularity curve with
infinitely strongly coupled perturbations. Thus one can consider that
the oscillatory solution and the superluminal configurations are described
by two separate EFT's. There are arguments \cite{Adams:2006sv} that
the presence of superluminality implies that the system cannot be
UV-completed in the usual local and Lorentz-invariant way. If the
superluminal configurations should be described by a disconnected
EFT different from the one which describes the limiting cycle and
neighboring configurations these arguments would not apply. 

Then in section (\ref{sec:Particular_Model}) we analyzed in details
the dynamics of the particular class of k-essence models from \cite{Bains:2015gpv}.
These theories have three free parameters. Following the spirit of
the Poincare-Bendixson theorem we analyzed for which values of the
parameters the limiting cycle is possible. This section serves for
the illustrative purposes of our analytical results from the previous
sections. We provide numerically obtained phase plots for a better
visualization. In this particular class of theories the configurations
with the superluminality are always screened by the singularity from
the time-crystal or limiting cycle. However, the latter is also disconnected
from the trivial Lorentz-invariant vacuum. Hence the EFT describing
the time-crystal does not have any Lorentz-invariant solutions. It
is interesting to understand under which conditions this feature allows
for the theory to have a Lorentz-invariant and local UV-completion. 

The models which can violate NEC and can cross the Phantom divide
without immediate pathologies are Generalized Galileons / Horndeski
theories \cite{Horndeski,Deffayet:2011gz,Kobayashi:2011nu}. Recently
it has been argued \cite{Libanov:2016kfc} that for the subclass of
these theories introduced in \cite{Deffayet:2010qz} it is still rather
hard to have a singularity free cosmological evolution without any
pathology and with an NEC-violating phase. Further these results were
generalized to whole class of Horndeski theories in \cite{Kobayashi:2016xpl}.
Both these works only consider external matter which does not have
any direct coupling to the considered Galileon scalar field. May be
a presence of an additional degree of freedom particularly coupled
to the Galileon can ameliorate the problem. 

To conclude, we find the idea of cosmological time-crystals rather
interesting, but it seems to be very hard to realize this idea in
a physically plausible way. \\

\acknowledgments The work of A.V. was supported by the J. E. Purkyn\v{e}
Fellowship of the Czech Academy of Sciences, by the Grant Agency of
the Czech Republic under the grant P201/12/G028 and by the International
Research Unit of Advanced Future Studies, Kyoto University Research
Coordination Alliance. Many of the results of the paper were obtained
during the visit of A.V. to the Yukawa Institute for Theoretical Physics,
Kyoto University. A.V. would like to express gratitude to the members
and staff of YITP for the warm hospitality. A.V. is also thankful
to the Galileo Galilei Institute for Theoretical Physics for the kind
hospitality and the INFN for partial support during the intermediate
stages of writing this paper. The paper was finally finished at the
Cargese Summer Institute: Quantum Gravity, Cosmology and Particle
Physics. A.V. is thankful to the organizers and staff for the kind
hospitality and partial financial support. 

\bibliographystyle{utcaps}
\addcontentsline{toc}{section}{\refname}\bibliography{Osc}

\end{document}